# Scattering matrix of arbitrary tight-binding Hamiltonians

C. Ramírez[*] and L.A. Medina-Amayo

Departamento de Física, Facultad de Ciencias, Universidad Nacional Autónoma de México, Apartado Postal 70-542, 04510 Ciudad de México, México

* Corresponding author: e-mail carlos@ciencias.unam.mx

**Abstract**
A novel efficient method to calculate the scattering matrix (SM) of arbitrary tight-binding Hamiltonians is proposed, including cases with multiterminal structures. In particular, the SM of two kind of fundamental structures are given, which can be used to obtain the SM of bigger systems iteratively. Also, a procedure to obtain the SM of layer-composed periodic leads is described. This method allows renormalization approaches, which permits computations over macroscopic length systems without introducing additional approximations. Finally, the transmission coefficient of a ring-shaped multiterminal system and the transmission function of a square-lattice nanoribbon with a reduced width region are calculated.

## 1. Introduction

The conductance is one of the most important properties of a material, which is also very susceptible to quantum effects caused by miniaturization. Theoretically, the conductance ($G$) for quantum systems in the nanoscale can be calculated from the Landauer formula [1]

$$G = \frac{e^2}{\pi \hbar} T \qquad (1)$$

where $T$ is the transmission coefficient for single-channel systems or the transmission function in multi-channel ones [2]. For Layered composed systems, fast techniques to determine the Landauer conductance based on the transfer matrix approach have been widely used [3-8], where the wavefunction behavior is iteratively obtained by means of multiplications of transfer matrixes. However, it is also well known that such multiplications could introduce important numerical instabilities during calculations [9]. To overcome these instabilities, calculations of the scattering matrix (SM) from the transfer matrix [10] or by wave function matching [11] have been proposed. SM of combined systems can then be found by using the Redheffer star product [12]. The SM relates the incoming and outgoing waves of a scattering region and in general also depends on the external systems. This implies the calculation of SM between every different layer interface, which could be computationally expensive. For electromagnetic waves, this issue is solved by separating layers by free space gaps of zero thickness [13], however in tight-binding Hamiltonian it is not immediately clear what structure can play the role of the free space gap.

In this article, we establish a novel efficient method to determine the SM of multiterminal systems described by tight binding Hamiltonians, showing that atomic chains can suitable substitute the mentioned free space gaps. In section II it is described the method that allow the calculation of the SM of a system in terms of those of two complementary subsystems. In section III the SM of two fundamental kinds of system are given, which permit the construction of any other system. In section IV a method to calculate the SM of layered leads is proposed. Finally, section V shows some results obtained through this method.

## 2. The Method

Let us consider a layer $A$ containing $N^A$ inside-sites and $M^A$ frontier-sites. The $n$-th frontier-site is connected to $P_n^A$ external periodic atomic-chains of infinite length with null self-energies and nearest-neighbor hopping integrals $t_C$, as exemplified in Figure 1(a). The Hamiltonian of this system can be written in terms of Wannier states as

$$\hat{H} = \sum_{n=1}^{N^A} \varepsilon_n^{AI} |I_n^A\rangle\langle I_n^A| + \sum_{n=1}^{M^A} \varepsilon_n^{AF} |F_n^A\rangle\langle F_n^A| + \sum_{n=1}^{N^A}\sum_{m=1}^{N^A} t_{n,m}^{AI} |I_n^A\rangle\langle I_m^A| + \sum_{n=1}^{M^A}\sum_{m=1}^{M^A} t_{n,m}^{AF} |F_n^A\rangle\langle F_m^A|$$
$$+ \sum_{n=1}^{M^A}\sum_{m=1}^{P_n^A}\sum_{l=1}^{\infty} t_C \left( |C_{n,m,l}^A\rangle\langle C_{n,m,l-1}^A| + |C_{n,m,l-1}^A\rangle\langle C_{n,m,l}^A| \right) + \sum_{n=1}^{N^A}\sum_{m=1}^{M^A} t_{n,m}^{AIF} \left( |I_n^A\rangle\langle F_m^A| + |F_m^A\rangle\langle I_n^A| \right)$$
, (2)

where $|I_n^A\rangle$, $|F_n^A\rangle$ and $|C_{n,m,l}^A\rangle$ are respectively Wannier states of inside-sites, frontier-sites, and sites in the external $m$-th chain attached to the $n$-th frontier site, $|C_{n,m,0}^A\rangle \equiv |F_n^A\rangle$, $\varepsilon_n^{AI}$ and $\varepsilon_n^{AF}$ are self-energies of inside- and frontier-sites, $t_{n,m}^{AI}$, $t_{n,m}^{AF}$ and $t_{n,m}^{AIF}$ are respectively hopping integrals between inside-inside, frontier-frontier and inside-frontier states with $t_{n,n}^{AI} = t_{n,n}^{AF} \equiv 0$. Eigenkets of Hamiltonian (2) can be written as

$$|\Psi_A\rangle = \sum_{n=1}^{N^A} a_n^I |I_n^A\rangle + \sum_{n=1}^{M^A} a_n^F |F_n^A\rangle + \sum_{n=1}^{M^A}\sum_{m=1}^{P_n^A}\sum_{l=1}^{\infty} a_{n,m,l}^C |C_{n,m,l}^A\rangle \qquad (3)$$

and for an energy $E$ the coefficients in Equation (3) accomplish

$$\left(\varepsilon_\alpha^{AI} - E\right)a_\alpha^I + \sum_{n=1}^{N^A} t_{\alpha,n}^{AI} a_n^I + \sum_{m=1}^{M^A} t_{\alpha,n}^{AIF} a_n^F = 0 , \qquad (4)$$

$$\left(\varepsilon_\beta^{AF} - E\right)a_\beta^F + \sum_{n=1}^{N^A} t_{n,\beta}^{AIF} a_n^I + \sum_{n=1}^{M^A} t_{\beta,n}^{AF} a_n^F + t_C \sum_{n=1}^{P_\beta^A} a_{\beta,n,1}^C = 0 \qquad (5)$$

and

$$-E a_{\beta,\gamma,\delta}^C + t_C \left( a_{\beta,\gamma,\delta-1}^C + a_{\beta,\gamma,\delta+1}^C \right) = 0 \qquad (6)$$

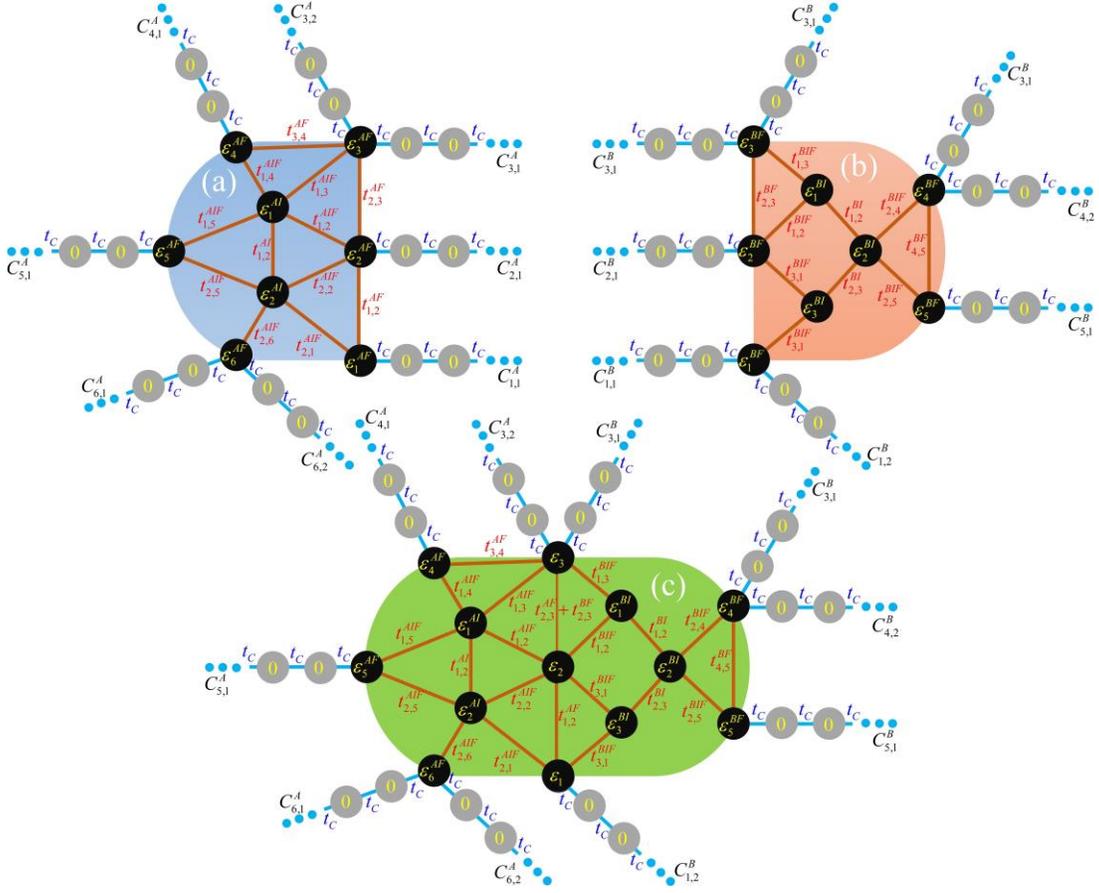

**FIG. 1.** Instances of (a) a layer $A$ with $N^A = 2$ inside-sites and $M^A = 6$ frontier-sites, (b) a layer $B$ with $N^B = 3$ inside-sites and $M^B = 5$ frontier-sites, (c) The combined layer $AB$ obtained when $A_{N,1}^{(+)} = B_{N,1}^{(-)}$ and $A_{N,1}^{(-)} = B_{N,1}^{(+)}$ being $N = 1, 2$ and $3$, where $\varepsilon_N = \varepsilon_N^{AF} + \varepsilon_N^{BF}$.

where $\alpha = 1, 2, \cdots, N^A$, $\beta = 1, 2, \cdots, M^A$, $\gamma = 1, 2, \cdots, P_\beta^A$, $\delta = 1, 2, 3, \cdots$ and $a_{\beta,\gamma,0}^C \equiv a_\beta^F$. The general solution of Equation (6) is

$$a_{\beta,\gamma,\delta}^C = A_{\beta,\gamma}^{(+)} e^{-i\kappa\delta} + A_{\beta,\gamma}^{(-)} e^{i\kappa\delta} \tag{7}$$

with

$$\kappa = \cos^{-1}\left(\frac{E}{2t_C}\right) \in [0, \pi]. \tag{8}$$

Hence $A_{\beta,\gamma}^{(+)}$ and $A_{\beta,\gamma}^{(-)}$ are respectively coefficients of incoming and outgoing waves of layer $A$. In particular we have

$$a_\beta^F = A_{\beta,1}^{(+)} + A_{\beta,1}^{(-)} \quad \text{and} \quad a_{\beta,1,1}^C = A_{\beta,1}^{(+)} e^{-i\kappa} + A_{\beta,1}^{(-)} e^{i\kappa}, \tag{9}$$

whose substitution in Equation (5) lead us to

$$\left(\varepsilon_\beta^{AF} - E\right)\left(A_{\beta,1}^{(+)} + A_{\beta,1}^{(-)}\right) + \sum_{n=1}^{N^A} t_{n,\beta}^{AIF} a_n^I + \sum_{n=1}^{M^A} t_{\beta,n}^{AF} a_n^F + t_C \sum_{n=2}^{P_\beta^A} a_{\beta,n,1}^C + t_C\left(A_{\beta,1}^{(+)} e^{-i\kappa} + A_{\beta,1}^{(-)} e^{i\kappa}\right) = 0. \tag{10}$$

Analogously, for another layer $B$ that contains $N^B$ inside-sites, $M^B$ frontier-sites and $P_n^B$ external chains attached to the $n$-th frontier site, solutions written as

$$|\Psi_B\rangle = \sum_{n=1}^{N^B} b_n^I |I_n^B\rangle + \sum_{n=1}^{M^B} b_n^F |F_n^B\rangle + \sum_{n=1}^{M^B}\sum_{m=1}^{P_n^B}\sum_{l=1}^{\infty} b_{n,m,l}^C |C_{n,m,l}^B\rangle \qquad (11)$$

lead us to equations

$$\left(\varepsilon_{\alpha'}^{BI} - E\right) b_{\alpha'}^I + \sum_{n=1}^{N^B} t_{\alpha',n}^{BI} b_n^I + \sum_{m=1}^{M^B} t_{\alpha',n}^{BIF} b_n^F = 0, \qquad (12)$$

$$-E b_{\beta',\gamma',\delta}^C + t_C \left(b_{\beta',\gamma',\delta-1}^C + b_{\beta',\gamma',\delta+1}^C\right) = 0, \qquad (13)$$

$$b_{\beta'}^F = B_{\beta',1}^{(+)} + B_{\beta',1}^{(-)} \quad \text{and} \quad b_{\beta',1,1}^C = B_{\beta',1}^{(+)} e^{-i\kappa} + B_{\beta',1}^{(-)} e^{i\kappa} \qquad (14)$$

$$\left(\varepsilon_{\beta'}^{BF} - E\right)\left(B_{\beta',1}^{(+)} + B_{\beta',1}^{(-)}\right) + \sum_{n=1}^{N^B} t_{n,\beta'}^{BIF} b_n^I + \sum_{n=1}^{M^B} t_{\beta',n}^{BF} b_n^F + t_C \sum_{n=2}^{P_{\beta'}^B} b_{\beta',n,1}^C + t_C \left(B_{\beta',1}^{(+)} e^{-i\kappa} + B_{\beta',1}^{(-)} e^{i\kappa}\right) = 0 \qquad (15)$$

where $\alpha' = 1, 2, \cdots, N^B$, $\beta' = 1, 2, \cdots, M^{B'}$, $\gamma' = 1, 2, \cdots, P_{\beta'}^B$, $\delta = 1, 2, 3, \cdots$ and $b_{\beta',\gamma',0}^C \equiv b_{\beta'}^F$.

Let us assume that

$$A_{N,1}^{(+)} = B_{N,1}^{(-)} \quad \text{and} \quad A_{N,1}^{(-)} = B_{N,1}^{(+)} \qquad (16)$$

for $N = 1, 2, 3, \cdots, M$, where $M \le \min(M^A, M^B)$. Then equations (9) and (14) imply that $a_N^F = b_N^F \equiv x_N$, and by summing equations (10) and (15) for $\beta = \beta' = N$ we obtain

$$\left(\varepsilon_N^{AF} + \varepsilon_N^{BF} - E\right) x_N + \sum_{n=1}^{N^A} t_{n,N}^{AIF} a_n^I + \sum_{n=1}^{N^B} t_{n,N}^{BIF} b_n^I + \sum_{n=M_A+1}^{M^A} t_{N,n}^{AF} a_n^F$$
$$+ \sum_{n=M_B+1}^{M^B} t_{N,n}^{BF} b_n^F + \sum_{N'=1}^{M} \left(t_{N,N'}^{AF} + t_{N,N'}^{BF}\right) x_{N'} + t_C \sum_{n=2}^{P_N^A} a_{N,n,1}^C + t_C \sum_{n=2}^{P_N^B} b_{N,n,1}^C = 0 \qquad (17)$$

where Equations (8), (9) and (14) have been used. On the other hand, Equations (4), (5), (6), (12), (13) and (15) for $\alpha = 1, 2, \cdots, N^A$, $\beta = N+1, N+2, \cdots, M^A$, $\gamma = 2, 3, \cdots, P_\beta^A$, $\alpha' = 1, 2, \cdots, N^B$, $\beta' = N+1, N+2, \cdots, M^B$ and $\gamma' = 2, 3, \cdots, P_{\beta'}^B$ only need to take into account that $a_N^F = b_N^F = x_N$ for $N = 1, 2, 3, \cdots, M$. These equations solve a combined layer $AB$ composed by the sites of layers $A$ and $B$, where there is a fusion between states $|F_N^A\rangle$ and $|F_N^B\rangle$ into a new one, $|x_N\rangle$, with self-energy $\varepsilon_N^{AF} + \varepsilon_N^{BF}$, hopping integrals between fused states given as $\langle x_{N'} | \hat{H} | x_N \rangle = t_{N,N'}^{AF} + t_{N,N'}^{BF}$ and connected to $P_N^A + P_N^B - 2$ external chains, while other self-energies, hopping integrals and connected chains remain unaltered, as exemplified in Figure 1(c).

Let us define $P^A$ as the number of external chains attached to layer $A$, i.e.,

$$P^A = \sum_{n=1}^{M^A} P_n^A \qquad (18)$$

Equations (4) to (7) allow us to write the coefficients of the outgoing waves as

$$\begin{pmatrix} \mathbf{A}_1^{(-)} \\ \mathbf{A}_2^{(-)} \end{pmatrix} = \mathbf{S}^A \begin{pmatrix} \mathbf{A}_1^{(+)} \\ \mathbf{A}_2^{(+)} \end{pmatrix} \equiv \begin{pmatrix} \mathbf{S}_{11}^A & \mathbf{S}_{12}^A \\ \mathbf{S}_{21}^A & \mathbf{S}_{22}^A \end{pmatrix} \begin{pmatrix} \mathbf{A}_1^{(+)} \\ \mathbf{A}_2^{(+)} \end{pmatrix} \tag{19}$$

where $\mathbf{S}^A$ is the SM of layer $A$, $\mathbf{S}_{11}^A$, $\mathbf{S}_{12}^A$, $\mathbf{S}_{21}^A$ and $\mathbf{S}_{22}^A$ are respectively $M \times M$, $M \times (P^A - M)$, $(P^A - M) \times M$ and $(P^A - M) \times (P^A - M)$ matrixes, $\mathbf{A}_1^{(\pm)}$ is a column vector of $M$ rows such as

$$\left( \mathbf{A}_1^{(\pm)} \right)_n = A_{n,1}^{(\pm)} \tag{20}$$

and $\mathbf{A}_2^{(\pm)}$ is a column vector of $P^A - M$ rows that contains the coefficients $A_{\beta,\gamma}^{(\pm)}$ that are not included in $\mathbf{A}_1^{(\pm)}$. Analogously, for layer $B$ we have

$$\begin{pmatrix} \mathbf{B}_1^{(-)} \\ \mathbf{B}_2^{(-)} \end{pmatrix} = \mathbf{S}^B \begin{pmatrix} \mathbf{B}_1^{(+)} \\ \mathbf{B}_2^{(+)} \end{pmatrix} \equiv \begin{pmatrix} \mathbf{S}_{11}^B & \mathbf{S}_{12}^B \\ \mathbf{S}_{21}^B & \mathbf{S}_{22}^B \end{pmatrix} \begin{pmatrix} \mathbf{B}_1^{(+)} \\ \mathbf{B}_2^{(+)} \end{pmatrix}. \tag{21}$$

Assumption (16) means $\mathbf{B}_1^{(+)} = \mathbf{A}_1^{(-)}$ and $\mathbf{B}_1^{(-)} = \mathbf{A}_1^{(+)}$, and together with equations (19) and (21) allow us to relate the incoming and outgoing waves of layer $AB$ as

$$\begin{pmatrix} \mathbf{A}_2^{(-)} \\ \mathbf{B}_2^{(-)} \end{pmatrix} = \begin{pmatrix} \mathbf{S}_{11}^{AB} & \mathbf{S}_{12}^{AB} \\ \mathbf{S}_{21}^{AB} & \mathbf{S}_{22}^{AB} \end{pmatrix} \begin{pmatrix} \mathbf{A}_2^{(+)} \\ \mathbf{B}_2^{(+)} \end{pmatrix} = \mathbf{S}^{AB} \begin{pmatrix} \mathbf{A}_2^{(+)} \\ \mathbf{B}_2^{(+)} \end{pmatrix} \tag{22}$$

where $\mathbf{S}^{AB}$ is the SM of the combined layer $AB$, while $\mathbf{S}_{11}^{AB}$, $\mathbf{S}_{12}^{AB}$, $\mathbf{S}_{21}^{AB}$ and $\mathbf{S}_{22}^{AB}$ are respectively $(P^A - M) \times (P^A - M)$, $(P^A - M) \times (P^B - M)$, $(P^B - M) \times (P^A - M)$ and $(P^B - M) \times (P^B - M)$ matrixes. As proved in Appendix A, they are given by

$$\begin{cases} \mathbf{S}_{11}^{AB} = \mathbf{S}_{22}^A + \mathbf{S}_{21}^A \left( \mathbf{I} - \mathbf{S}_{11}^B \mathbf{S}_{11}^A \right)^{-1} \mathbf{S}_{11}^B \mathbf{S}_{12}^A \\ \mathbf{S}_{12}^{AB} = \mathbf{S}_{21}^A \left( \mathbf{I} - \mathbf{S}_{11}^B \mathbf{S}_{11}^A \right)^{-1} \mathbf{S}_{12}^B \\ \mathbf{S}_{21}^{AB} = \mathbf{S}_{21}^B \left( \mathbf{I} - \mathbf{S}_{11}^A \mathbf{S}_{11}^B \right)^{-1} \mathbf{S}_{12}^A \\ \mathbf{S}_{22}^{AB} = \mathbf{S}_{22}^B + \mathbf{S}_{21}^B \left( \mathbf{I} - \mathbf{S}_{11}^A \mathbf{S}_{11}^B \right)^{-1} \mathbf{S}_{11}^A \mathbf{S}_{12}^B \end{cases}, \tag{23}$$

where $\mathbf{I}$ is the $M \times M$ identity matrix. It is worth to mention that if we would have interchange the position of $\mathbf{A}_1^{(\pm)}$ and $\mathbf{A}_2^{(\pm)}$ in Equation (19), the matrix $\mathbf{S}^{AB}$ could have been obtained through the Redheffer star product [13].

In resume, the method described in this section allow us to obtain the SM of a system, in terms of the SM of its components. In consequence, by iterating this method, we can obtain the SM of bigger systems.

## 3. The Building Blocks

There are two kind of fundamental layers that allow us to determine the SM of layers with Hamiltonian (2) through the method described in the previous section, the site- and the bond-layers, which are described below

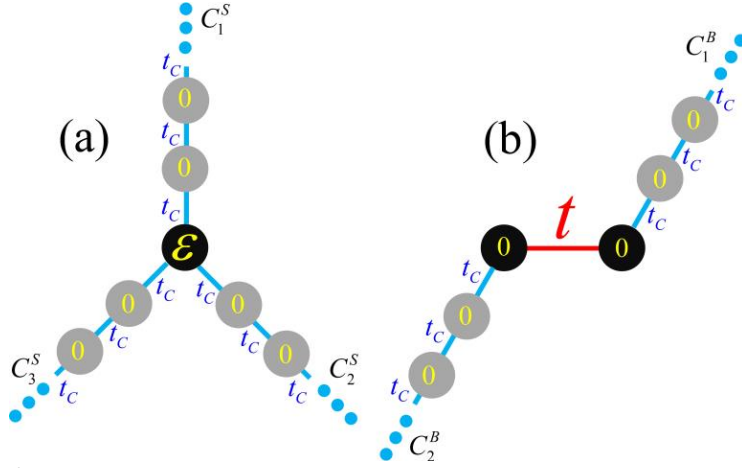

**FIG. 2.** Schematic representations of (a) a site layer with self-energy $\varepsilon$ and $P=3$ coupled chains and (b) a bond layer with hopping integral $t$.

### 3.1. Site-layer

A site-layer contains only one frontier site with self-energy $\varepsilon$ and $P$ external chains attached to it, as shown in Fig. 2(a). Straightforward calculation give us the $P \times P$ scattering matrix of this layer as

$$\left(\mathbf{S}^{site}\right)_{nm} = \frac{2it_C \sin\kappa}{\varepsilon - E + t_C e^{i\kappa} P} - \delta_{nm}, \quad (24)$$

where $\kappa$ is given by Equation (8).

Site-layers SM with $\varepsilon = 0$ and $P > 2$ can be used to increase the number of external chains coupled to a frontier site, while those with $P = 1$ remove an external chain. If $\varepsilon \neq 0$ and $P = 2$, such SM can be employed to increase by $\varepsilon$ the self-energy of a frontier site while keeping the number of external chains attached to it.

### 3.2. Bond-layer

A bond-layer has two frontier sites, with one external chain attached to each frontier site, being $t \neq 0$ the hopping integral between such frontier sites, as shown in Fig. 2(b). This system has a $2 \times 2$ scattering matrix given by

$$\left(\mathbf{S}^{site}\right)_{nm} = \begin{cases} r & \text{if } n = m \\ \frac{t_C}{t}\left(e^{i\kappa} + re^{-i\kappa}\right) & \text{if } n \neq m \end{cases}, \quad (25)$$

where

$$r = -\frac{t^2 - t_C^2}{t^2 - t_C^2 e^{-2i\kappa}}. \quad (26)$$

Bond-layers are then used to connect site layers. In general, a bond layer permit us to augment the number of sites in an existing layer, and in combination with site-layers, allow us to treat any general system with Hamiltonian (2).

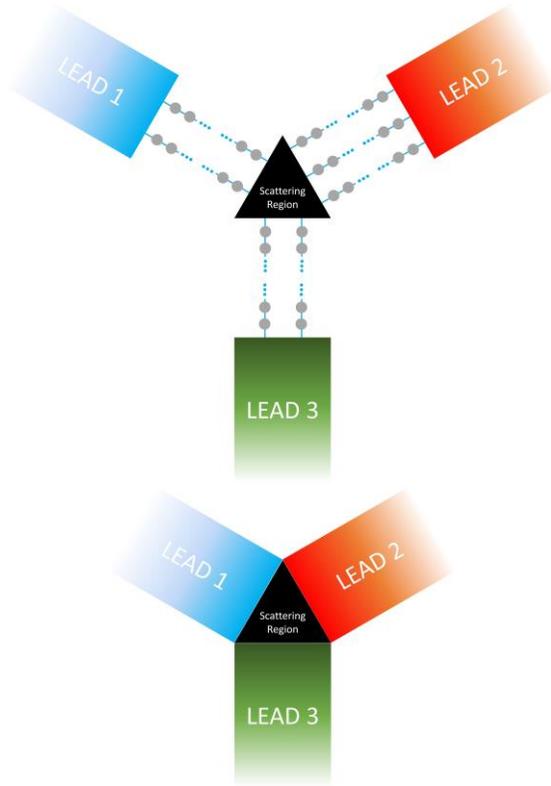

**FIG. 3.** Process of obtaining the scattering matrix of a general system, by using the method described in Section 2. In the showed instance, a scattering region is connected to three semi-infinite leads.

## 4. The Leads

The general scattering problem consists of a scattering region connected to semi-infinite quantum leads, which have well defined incoming and outgoing modes [2]. In this sense, we need the scattering matrixes between the leads and external atomic-chains, in order to be able to connect the scattered region to such leads, as illustrated in Fig. 3. Notice that there are not external chains in the final system, then the incoming and outgoing waves considered in the resulting scattering matrix are merely those of the leads.

Let us consider a periodic lead, as shown in Fig. 4, where a transversal layer of $Q$ bonded sites is repeated longitudinally, by connecting each site to their equivalent ones in contiguous layers by hopping integrals $t_L$. The last transversal layer (blue balls) are the frontier sites of the lead, which have null self-energies and null hopping integrals between them.

Let us suppose that the eigenvalues of the transversal layer (top of Figure 4) are

$$E_T^{(1)}, E_T^{(2)}, \cdots, E_T^{(Q)} \tag{27}$$

which are ordered such as $\left|(E - E_T^{(s)})/2t_L\right| < 1$ for $s = 1, 2, \cdots, \tilde{Q}$ and $\left|(E - E_T^{(s)})/2t_L\right| \geq 1$ for $s = \tilde{Q}+1, \cdots, Q$ for a given energy $E$, with corresponding orthonormalized eigenkets

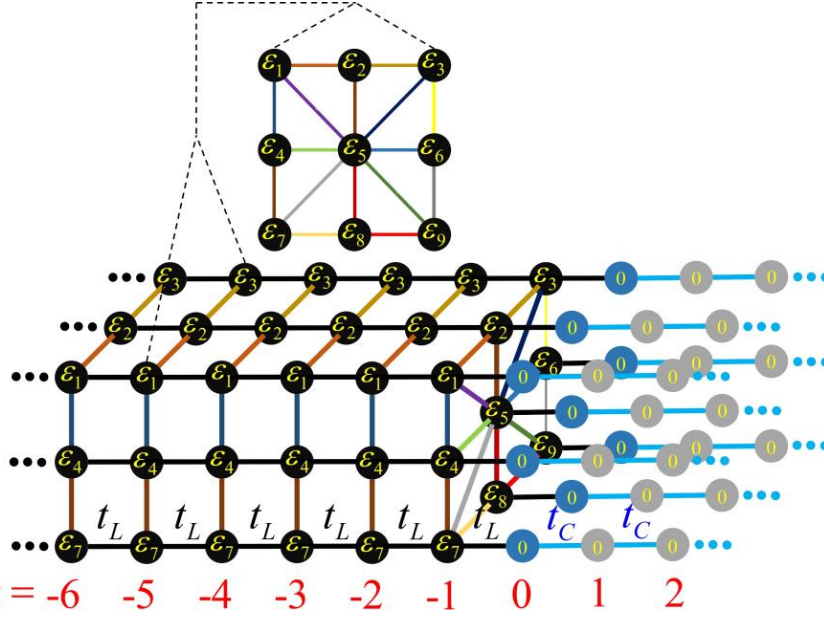

**FIG. 4.** A periodic lead composed by transversal layers of $Q=9$ sites, where $n$ labels such layers. In particular, we find the frontier sites of the lead at $n=0$ (blue balls), inside-sites of lead for $n<0$ (black balls) and sites of external atomic-chains when $n>0$ (gray balls). In the top of figure, it is shown the transversal layer used to construct the lead.

$$\left|\Psi_T^{(s)}\right\rangle = \sum_{m=1}^{Q} \alpha_m^{(s)} \left|\varepsilon_m\right\rangle \tag{28}$$

where $s=1,2,\cdots,Q$ and $\left|\varepsilon_m\right\rangle$ is the Wannier state with self-energy $\varepsilon_m$.

Eigenkets of the system in Fig. 4 can be written as

$$\left|\Psi_{Lead}\right\rangle = \sum_{n=-\infty}^{\infty} \sum_{m=1}^{Q} C_{n,m} \left|n,m\right\rangle \tag{29}$$

being $\left|n,m\right\rangle$ the Wannier state of the $m$-th site of the $n$-th transversal layer. It follows that those coefficients with $n=0,-1,-2,\cdots$ can be written as

$$C_{n,m} = \sum_{s=1}^{\tilde{Q}} \alpha_m^{(s)} \left(L_s^{(+)} e^{in\kappa_s} + L_s^{(-)} e^{-in\kappa_s}\right) + \sum_{s=\tilde{Q}+1}^{Q} \alpha_m^{(s)} L_s e^{n\kappa_s} F_s^n \tag{30}$$

where $L_s^{(+)}$ and $L_s^{(-)}$ are respectively amplitudes of the incoming and outgoing wavefunctions of an open-channel $s$ $\left(\text{where } \left|(E-E_T^{(s)})/2t_L\right|<1\right)$, $L_s$ is the coefficient of an evanescent wavefunction of a close-channel $s$ $\left(\text{where } \left|(E-E_T^{(s)})/2t_L\right|\geq 1\right)$, $\tilde{Q}=\tilde{Q}(E)$ is the number of open channels for a given energy $E$,

$$F_s = \begin{cases} 1 & \text{if } (E-E_T^{(s)})/2t_L \geq 1 \\ -1 & \text{if } (E-E_T^{(s)})/2t_L \leq -1 \end{cases}, \tag{31}$$

while $\kappa_s$ is

$$\kappa_s = \begin{cases} \cos^{-1}\left(\dfrac{E - E_T^{(s)}}{2t_L}\right) \in (0, \pi] & \text{if } s = 1, 2, \cdots, \tilde{Q} \\ \cosh^{-1}\left|\dfrac{E - E_T^{(s)}}{2t_L}\right| \in [0, \infty) & \text{if } s = \tilde{Q} + 1, \cdots, Q \end{cases}. \tag{32}$$

On the other hand equation (7) indicates that

$$C_{n,m} = A_m^{(+)} e^{-in\kappa} + A_m^{(-)} e^{in\kappa} \tag{33}$$

for $n = 0, 1, 2, \cdots$, where $A_m^{(+)}$ and $A_m^{(-)}$ are respectively the coefficients of the incoming and outgoing waves of the external chain attached to the $m$-th frontier site, $\kappa$ is given by equation (8) and $t_C$ is big enough to maintain open channels in the external chains, $i.e.$, $|E/2t_C| < 1$. Moreover, we have

$$-E C_{0,m} + t_L C_{-1,m} + t_C C_{1,m} = 0. \tag{34}$$

Equations (30), (33) and (34) imply

$$\begin{pmatrix} \mathbf{C}_0 \\ \mathbf{C}_{-1} \end{pmatrix} = \begin{pmatrix} \mathbf{M}_{1,1}^L & \mathbf{M}_{1,2}^L & \mathbf{M}_{1,3}^L \\ \mathbf{M}_{2,1}^L & \mathbf{M}_{2,2}^L & \mathbf{M}_{2,3}^L \end{pmatrix} \begin{pmatrix} \mathbf{L}^{(+)} \\ \mathbf{L}^{(-)} \\ \mathbf{L} \end{pmatrix} = \begin{pmatrix} \mathbf{M}_{1,1}^A & \mathbf{M}_{1,2}^A \\ \mathbf{M}_{2,1}^A & \mathbf{M}_{2,2}^A \end{pmatrix} \begin{pmatrix} \mathbf{A}^{(+)} \\ \mathbf{A}^{(-)} \end{pmatrix} \tag{35}$$

where $\mathbf{C}_n$ and $\mathbf{A}^{(\pm)}$ are column vectors of $Q$ rows such as $(\mathbf{C}_n)_m = C_{n,m}$ and $(\mathbf{A}^{(\pm)})_n = A_n^{(\pm)}$; $\mathbf{L}^{(\pm)}$ are column vectors of $\tilde{Q}$ rows with $(\mathbf{L}^{(\pm)})_m = L_m^{(\pm)}$; $\mathbf{L}$ is a column vector of $Q - \tilde{Q}$ rows with $(\mathbf{L})_m = L_{m+\tilde{Q}}$; $\mathbf{M}_{1,1}^L$, $\mathbf{M}_{1,2}^L$, $\mathbf{M}_{2,1}^L$, $\mathbf{M}_{2,2}^L$ are $Q \times \tilde{Q}$ matrixes given by

$$\left(\mathbf{M}_{1,1}^L\right)_{n,m} = \left(\mathbf{M}_{1,2}^L\right)_{n,m} = e^{i\kappa_m} \left(\mathbf{M}_{2,1}^L\right)_{n,m} = e^{-i\kappa_m} \left(\mathbf{M}_{2,2}^L\right)_{n,m} = \alpha_n^{(m)}; \tag{36}$$

$\mathbf{M}_{1,3}^L$, $\mathbf{M}_{2,3}^L$ are $Q \times (Q - \tilde{Q})$ matrixes which accomplish

$$\left(\mathbf{M}_{1,3}^L\right)_{n,m} = F_s e^{\kappa_{m+\tilde{Q}}} \left(\mathbf{M}_{2,3}^L\right)_{n,m} = \alpha_n^{(m+\tilde{Q})}; \tag{37}$$

and $\mathbf{M}_{i,j}^A$ are $Q \times Q$ matrixes written as

$$\left(\mathbf{M}_{1,1}^A\right)_{n,m} = \left(\mathbf{M}_{1,2}^A\right)_{n,m} = \delta_{n,m}, \tag{38}$$

$$\left(\mathbf{M}_{2,1}^A\right)_{n,m} = \frac{1}{t_L}\left(E - t_C e^{-i\kappa}\right)\delta_{n,m} \tag{39}$$

and

$$\left(\mathbf{M}_{2,2}^A\right)_{n,m} = \frac{1}{t_L}\left(E - t_C e^{i\kappa}\right)\delta_{n,m}. \tag{40}$$

From equation (35) we obtain

$$\begin{pmatrix} \mathbf{A}^{(-)} \\ \mathbf{L}^{(-)} \\ \mathbf{L} \end{pmatrix} = \begin{pmatrix} \mathbf{C}_{1,1}^L & \mathbf{C}_{1,2}^L \\ \mathbf{C}_{2,1}^L & \mathbf{C}_{2,2}^L \end{pmatrix} \begin{pmatrix} \mathbf{A}^{(+)} \\ \mathbf{L}^{(+)} \end{pmatrix} \tag{41}$$

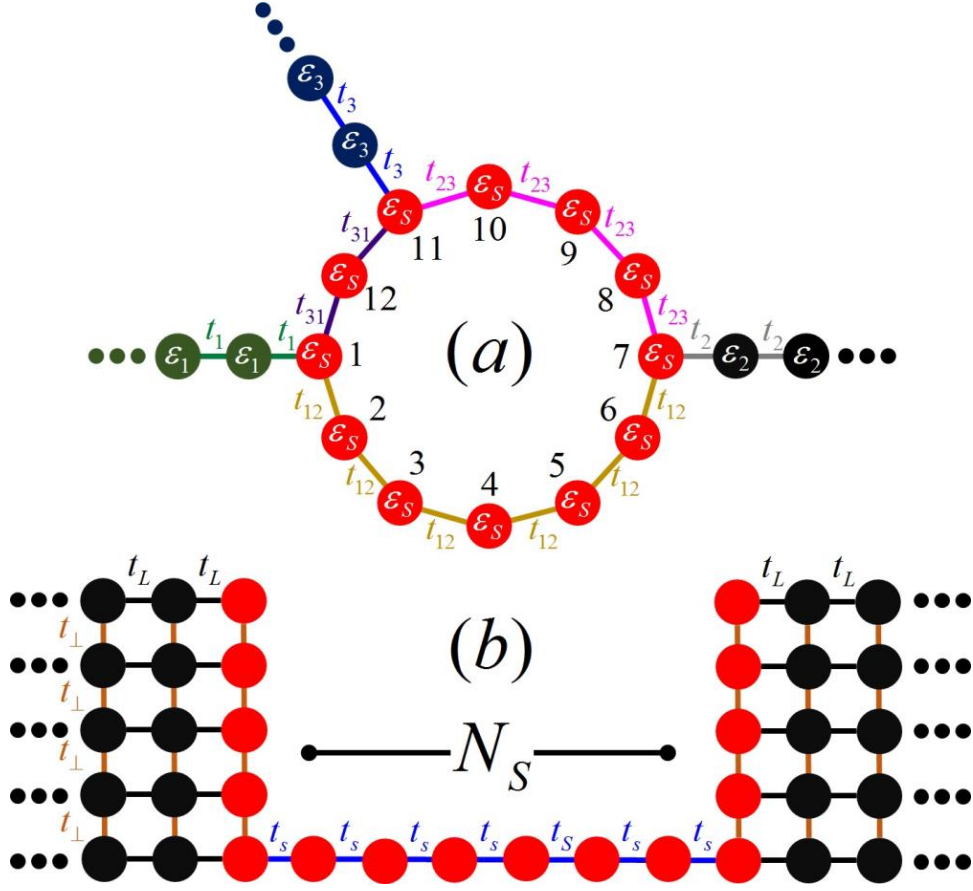

**FIG. 5.** (a) A ring-shaped system of $N_S = 12$ sites connected to $N_T = 3$ periodic chains in sites $T_1 = 1$, $T_2 = 7$ and $T_3 = 11$, with hopping integrals and self-energies indicated in figure. (b) A square-lattice nanoribbon with $M_T = 5$ atoms of width, is reduced to one of $M_S = 1$ atoms and then returned to its original width, having a reduced-width region of $N_S = 6$ atoms of length.

where as demonstrated in Appendix B,

$$\begin{cases} \mathbf{C}_{1,1}^L = \left[ \mathbf{M}_{2,2}^A - \tilde{\mathbf{M}}_2^L \left( \tilde{\mathbf{M}}_1^L \right)^{-1} \mathbf{M}_{1,2}^A \right]^{-1} \left[ \tilde{\mathbf{M}}_2^L \left( \tilde{\mathbf{M}}_1^L \right)^{-1} \mathbf{M}_{1,1}^A - \mathbf{M}_{2,1}^A \right] \\ \mathbf{C}_{1,2}^L = \left[ \mathbf{M}_{2,2}^A - \tilde{\mathbf{M}}_2^L \left( \tilde{\mathbf{M}}_1^L \right)^{-1} \mathbf{M}_{1,2}^A \right]^{-1} \left[ \mathbf{M}_{2,1}^L - \tilde{\mathbf{M}}_2^L \left( \tilde{\mathbf{M}}_1^L \right)^{-1} \mathbf{M}_{1,1}^L \right] \\ \mathbf{C}_{2,1}^L = \left[ \tilde{\mathbf{M}}_2^L - \mathbf{M}_{2,2}^A \left( \mathbf{M}_{1,2}^A \right)^{-1} \tilde{\mathbf{M}}_1^L \right]^{-1} \left[ \mathbf{M}_{2,1}^A - \mathbf{M}_{2,2}^A \left( \mathbf{M}_{1,2}^A \right)^{-1} \mathbf{M}_{1,1}^A \right] \\ \mathbf{C}_{2,2}^L = \left[ \tilde{\mathbf{M}}_2^L - \mathbf{M}_{2,2}^A \left( \mathbf{M}_{1,2}^A \right)^{-1} \tilde{\mathbf{M}}_1^L \right]^{-1} \left[ \mathbf{M}_{2,2}^A \left( \mathbf{M}_{1,2}^A \right)^{-1} \mathbf{M}_{1,1}^L - \mathbf{M}_{2,1}^L \right] \end{cases}, \quad (42)$$

being $\tilde{\mathbf{M}}_i^L \equiv \begin{pmatrix} \mathbf{M}_{i,2}^L & \mathbf{M}_{i,3}^L \end{pmatrix}$. In particular, Equation (41) allow us to write

$$\begin{pmatrix} \mathbf{A}^{(-)} \\ \mathbf{L}^{(-)} \end{pmatrix} = \mathbf{S}^L \begin{pmatrix} \mathbf{A}^{(+)} \\ \mathbf{L}^{(+)} \end{pmatrix}. \quad (43)$$

Hence the scattering matrix of the lead, $\mathbf{S}^L$, correspond to the first $Q+\tilde{Q}$ rows of the matrix in Equation (41).

## 5. Results and Discussion

In order to illustrate the usefulness of this method, let us consider the systems shown in Figure 5, corresponding to (a) a ring-shape multiterminal system and (b) a square lattice nanoribbon with a reduced-width region.

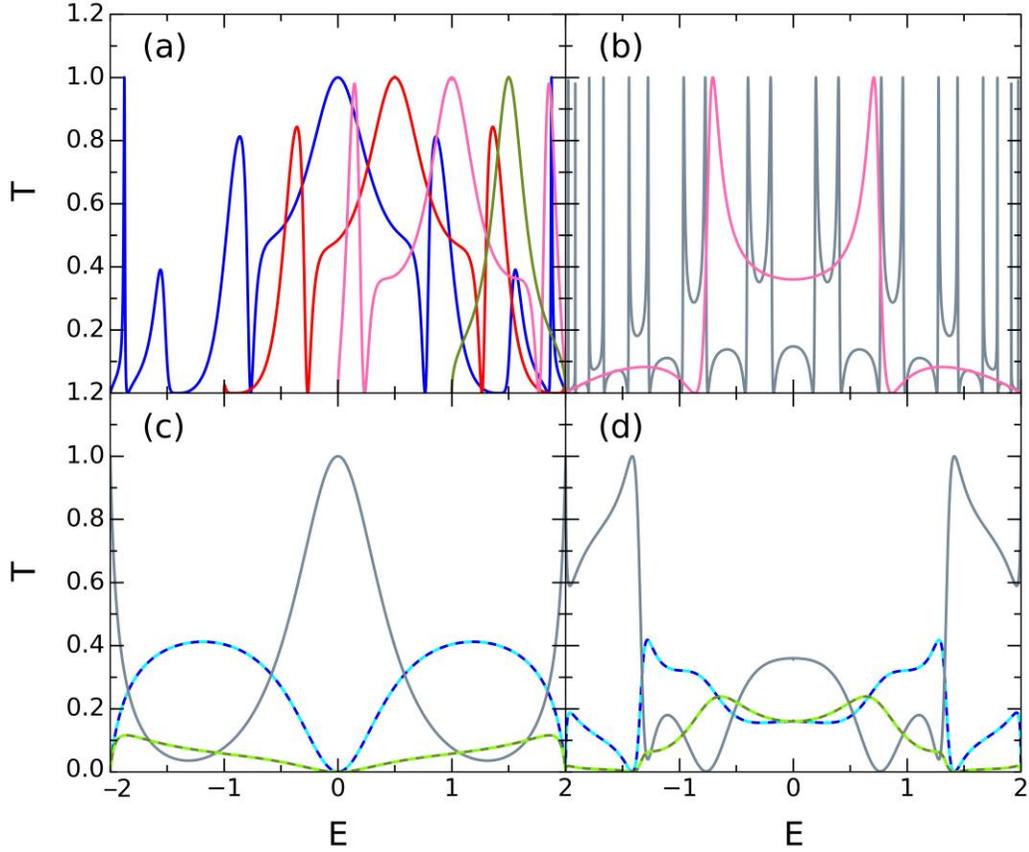

**FIG 6.** Transmission coefficient ($T$) of the system in Figure 5(a) for the cases of (a) two terminals with $N_S = 16$, $T_1 = 1$, $T_2 = 7$, $t_{i,j} = t_i = t$, $\varepsilon_2 = 0$, $\varepsilon_S = (\varepsilon_1 + \varepsilon_2)/2$ and $\varepsilon_1 = 0$ (blue line), $\varepsilon_1 = |t|$ (red line), $\varepsilon_1 = 2|t|$ (pink line) and $\varepsilon_1 = 3|t|$ (green line); (b) two terminals with $T_1 = 1$, $T_2 = 2$, $t_{21} = t_i = t$, $\varepsilon_i = \varepsilon_S = 0$, $t_{12} = 4t$ and $N_S = 4$ (pink line) and $N_S = 22$ (gray line); $N_T = 5$ equidistant terminals, being $t_{ij} = t_i = t$, $\varepsilon_i = \varepsilon_S = 0$, showing the transmittance between contiguous terminals (blue dashed lines), between nonadjacent terminals (green dashed lines) and the reflectance (gray solid lines) for (c) $N_S = 10$ and (d) $N_S = 20$. Energy is measured in units of $t$.

In Fig. 6 it is shown the transmission coefficient between terminals for the system of Fig. 5(a) when (a) there are two terminals with $N_S = 16$, $T_1 = 1$, $T_2 = 7$, $t_{i,j} = t_i = t$, $\varepsilon_1 = 0$, $\varepsilon_S = (\varepsilon_1 + \varepsilon_2)/2$ and $\varepsilon_1/|t| = 0,1,2,3$; (b) there are two terminals with $T_1 = 1$, $T_2 = 2$, $t_{21} = t_i = t$, $\varepsilon_i = \varepsilon_S = 0$, $t_{12} = 4t$ and $N_S = 4$ or $22$; there are $N_T = 5$ equidistant terminals, being $t_{ij} = t_i = t$, $\varepsilon_i = \varepsilon_S = 0$ and (c) $N_S = 10$ or (d) $N_S = 20$. The results obtained in Fig. 6(a) and 6(b) are in agreement with those respectively found in Fig. 3 of Ref. [14], and Fig. 3 and 4 of Ref. [15]. On the other hand, for the multiterminal systems of Fig 6(c) and 6(d), notice that the transmittance is greater between contiguous terminals in comparison to that between nonadjacent terminals, except around zero energy in Fig. 6(d), where both transmittances are of the same order. In appendix C, it is shown that for a system with five equidistant terminals and $N_S = 10n$, being $n = 1, 2, 3, \cdots$, there are total reflectance states for energies

$$E = 2t \cos\left(\frac{m\pi}{2n}\right) \qquad (44)$$

with $m = 1, 3, 5, \cdots, 2n-1$. These total reflection states are found at $E = 0$ in Fig. 6(c) and at $E = \pm\sqrt{2}t$ in Fig. 6(d).

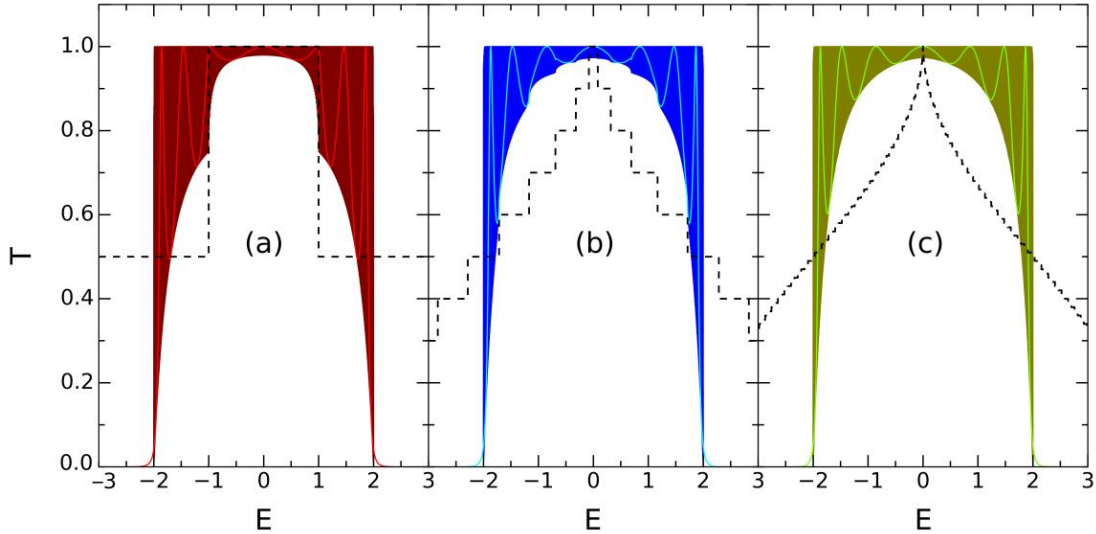

**FIG 7.** Transmission function ($T$) as a function of energy ($E$) for the system in Fig. 5(b) with hopping integrals $t$, null self-energies, and parameters (a) $M_T = 2$, $M_S = 1$ and $N_S = 14$ (red lines) or $N_S = 1022$ (wine lines); (b) $M_T = 10$, $M_S = 1$ and $N_S = 14$ (cyan lines) or $N_S = 1022$ (blue lines); and (c) $M_T = 100$, $M_S = 1$ and $N_S = 14$ (green lines) or $N_S = 1022$ (dark yellow lines). It is also shown $T/M_T$ corresponding the infinite square-lattice nanoribbon without defects (black dotted lines), where $T$ becomes equal to the number of open channels in the leads. Energy is measured in units of $t$.

Figure 7 shows the effective transmission function ($T$), defined as the summation of the transmittances from the left to the right lead [2], for the system in Fig. 5(b) when all self-energies are zero and hopping integrals are $t$, (a) $M_T = 2$, $M_S = 1$ and $N_S = 14$ or $1022$; (b) $M_T = 10$, $M_S = 1$ and $N_S = 14$ or $1022$; and (c) $M_T = 100$, $M_S = 1$ and $N_S = 14$ or $1022$. Notice that in all cases leads are connected by an atomic-chain that have extended states only if $|E| < 2|t|$, which establish a superior envelope that is essentially unaffected by the width of leads. In fact, for the cases with $N_S = 14$ atoms, we can still find some non-zero transmittance states just above (below) the energy $E = 2|t|$ ($-2|t|$), caused by tunneling, which vanish when $N_S$ grows. On the other hand, there is an inferior envelope which strongly depends on the width of leads. Actually, we can notice abrupt changes in this envelope when the number of open channels in the leads changes (black dotted lines). Finally, a greater $N_S$ increases the oscillations maintaining the mentioned envelopes.

## 6. Conclusions

An efficient method to calculate the scattering matrix of arbitrary tight-binding Hamiltonians with multiterminal atomic-chain leads has been discussed. We also develop the extension to cases with layer-composed leads, whose elements are connected by interlayer hopping integrals $t_L$, as schematically shown in Fig. 4. In particular, we give the scattering matrixes of site and bond layers, which are the fundamental elements that allow determining the scattering matrix of bigger systems. In order to validate the correctness of this method, we reproduce in Fig. 6(a) and 6(b) some results respectively found in Ref. [14] and [15], and obtain the transmittances for a multiterminal system in Fig. 6(c) and 6(d), whose total reflectance states are determined analytically in Appendix C. On the other hand, the transmission function is calculated for a square-lattice nanoribbon with a region of reduced width, finding that its value depends on the number of open-channels in the leads, and the length and width of the reduced-width region. Finally, it is worth to mention that this method is compatible with renormalization techniques, which allow calculations for macroscopic length systems, for example, by using the generalized Fibonacci building rule [8] or the doubling algorithm [13]. Moreover, since the scattering matrix is unitary, this method reduces numerical error in comparison to transfer-matrix approaches [10].

## Acknowledgments

Computations were performed at Miztli under project SC16-1-S-80 of DGTIC-UNAM.

## Appendix A

In this appendix it is proved that given equations (19), (21) and (22) with $\mathbf{B}_1^{(+)} = \mathbf{A}_1^{(-)}$ and $\mathbf{B}_1^{(-)} = \mathbf{A}_1^{(+)}$ it is obtained Equation (23).

Equations (19) and (21) define the next system of equations

$$\mathbf{C} = \mathbf{S}_{11}^A \mathbf{D} + \mathbf{S}_{12}^A \mathbf{A}_2^{(+)} \tag{A.1}$$

$$\mathbf{A}_2^{(-)} = \mathbf{S}_{21}^A \mathbf{D} + \mathbf{S}_{22}^A \mathbf{A}_2^{(+)} \tag{A.2}$$

$$\mathbf{D} = \mathbf{S}_{11}^B \mathbf{C} + \mathbf{S}_{12}^B \mathbf{B}_2^{(+)} \tag{A.3}$$

$$\mathbf{B}_2^{(-)} = \mathbf{S}_{21}^B \mathbf{C} + \mathbf{S}_{22}^B \mathbf{B}_2^{(+)} \tag{A.4}$$

where $\mathbf{C} \equiv \mathbf{B}_1^{(+)} = \mathbf{A}_1^{(-)}$ and $\mathbf{D} \equiv \mathbf{B}_1^{(-)} = \mathbf{A}_1^{(+)}$. Substituting equation (A.3) into (A.1) lead us to

$$\mathbf{C} = \left(\mathbf{I} - \mathbf{S}_{11}^A \mathbf{S}_{11}^B\right)^{-1} \mathbf{S}_{12}^A \mathbf{A}_2^{(+)} + \left(\mathbf{I} - \mathbf{S}_{11}^A \mathbf{S}_{11}^B\right)^{-1} \mathbf{S}_{11}^A \mathbf{S}_{12}^B \mathbf{B}_2^{(+)}, \tag{A.5}$$

while substitution of equation (A.1) into (A.3) imply

$$\mathbf{D} = \left(\mathbf{I} - \mathbf{S}_{11}^B \mathbf{S}_{11}^A\right)^{-1} \mathbf{S}_{11}^B \mathbf{S}_{12}^A \mathbf{A}_2^{(+)} + \left(\mathbf{I} - \mathbf{S}_{11}^B \mathbf{S}_{11}^A\right)^{-1} \mathbf{S}_{12}^B \mathbf{B}_2^{(+)}. \tag{A.6}$$

Equations (A.2) and (A.6) give us

$$\mathbf{A}_2^{(-)} = \left[\mathbf{S}_{22}^A + \mathbf{S}_{21}^A \left(\mathbf{I} - \mathbf{S}_{11}^B \mathbf{S}_{11}^A\right)^{-1} \mathbf{S}_{11}^B \mathbf{S}_{12}^A\right] \mathbf{A}_2^{(+)} + \mathbf{S}_{21}^A \left(\mathbf{I} - \mathbf{S}_{11}^B \mathbf{S}_{11}^A\right)^{-1} \mathbf{S}_{12}^B \mathbf{B}_2^{(+)}, \tag{A.7}$$

and finally from Equations (A.4) and (A.5) we have

$$\mathbf{B}_2^{(-)} = \mathbf{S}_{21}^B \left(\mathbf{I} - \mathbf{S}_{11}^A \mathbf{S}_{11}^B\right)^{-1} \mathbf{S}_{12}^A \mathbf{A}_2^{(+)} + \left[\mathbf{S}_{22}^B + \mathbf{S}_{21}^B \left(\mathbf{I} - \mathbf{S}_{11}^A \mathbf{S}_{11}^B\right)^{-1} \mathbf{S}_{11}^A \mathbf{S}_{12}^B\right] \mathbf{B}_2^{(+)}. \tag{A.8}$$

From Equations (A.7), (A.8) and (22) we obtain equation (23).

## Appendix B

In this appendix it is proved Equation (42) from Equation (35).

Starting from Equation (35) we have

$$\mathbf{M}_{1,1}^L \mathbf{L}^{(+)} + \tilde{\mathbf{M}}_1^L \begin{pmatrix} \mathbf{L}^{(-)} \\ \mathbf{L} \end{pmatrix} = \mathbf{M}_{1,1}^A \mathbf{A}^{(+)} + \mathbf{M}_{1,2}^A \mathbf{A}^{(-)} \tag{B.9}$$

and

$$\mathbf{M}_{2,1}^L \mathbf{L}^{(+)} + \tilde{\mathbf{M}}_2^L \begin{pmatrix} \mathbf{L}^{(-)} \\ \mathbf{L} \end{pmatrix} = \mathbf{M}_{2,1}^A \mathbf{A}^{(+)} + \mathbf{M}_{2,2}^A \mathbf{A}^{(-)}, \tag{B.10}$$

where $\tilde{\mathbf{M}}_i^L \equiv \begin{pmatrix} \mathbf{M}_{i,2}^L & \mathbf{M}_{i,3}^L \end{pmatrix}$ is a $Q \times Q$ matrix. From Equation (B.9) we obtain

$$\begin{pmatrix} \mathbf{L}^{(-)} \\ \mathbf{L} \end{pmatrix} = \left(\tilde{\mathbf{M}}_1^L\right)^{-1} \left[\mathbf{M}_{1,1}^A \mathbf{A}^{(+)} - \mathbf{M}_{1,1}^L \mathbf{L}^{(+)} + \mathbf{M}_{1,2}^A \mathbf{A}^{(-)}\right] \tag{B.11}$$

and

$$\mathbf{A}^{(-)} = \left(\mathbf{M}_{1,2}^A\right)^{-1} \left[\mathbf{M}_{1,1}^L \mathbf{L}^{(+)} - \mathbf{M}_{1,1}^A \mathbf{A}^{(+)} + \tilde{\mathbf{M}}_1^L \begin{pmatrix} \mathbf{L}^{(-)} \\ \mathbf{L} \end{pmatrix}\right]. \tag{B.12}$$

Substituting Equation (B.11) in (B.10) lead us to

$$\mathbf{A}^{(-)} = \left[\mathbf{M}_{2,2}^A - \tilde{\mathbf{M}}_2^L \left(\tilde{\mathbf{M}}_1^L\right)^{-1} \mathbf{M}_{1,2}^A\right]^{-1} \left[\tilde{\mathbf{M}}_2^L \left(\tilde{\mathbf{M}}_1^L\right)^{-1} \mathbf{M}_{1,1}^A - \mathbf{M}_{2,1}^A\right] \mathbf{A}^{(+)}$$
$$+ \left[\mathbf{M}_{2,2}^A - \tilde{\mathbf{M}}_2^L \left(\tilde{\mathbf{M}}_1^L\right)^{-1} \mathbf{M}_{1,2}^A\right]^{-1} \left[\mathbf{M}_{2,1}^L - \tilde{\mathbf{M}}_2^L \left(\tilde{\mathbf{M}}_1^L\right)^{-1} \mathbf{M}_{1,1}^L\right] \mathbf{L}^{(+)} \tag{B.13}$$

On the other hand, substituting Equation (B.12) in (B.10) give us

$$\begin{pmatrix} \mathbf{L}^{(-)} \\ \mathbf{L} \end{pmatrix} = \left[ \tilde{\mathbf{M}}_2^L - \mathbf{M}_{2,2}^A \left( \mathbf{M}_{1,2}^A \right)^{-1} \tilde{\mathbf{M}}_1^L \right]^{-1} \left[ \mathbf{M}_{2,1}^A - \mathbf{M}_{2,2}^A \left( \mathbf{M}_{1,2}^A \right)^{-1} \mathbf{M}_{1,1}^A \right] \mathbf{A}^{(+)}$$
$$+ \left[ \tilde{\mathbf{M}}_2^L - \mathbf{M}_{2,2}^A \left( \mathbf{M}_{1,2}^A \right)^{-1} \tilde{\mathbf{M}}_1^L \right]^{-1} \left[ \mathbf{M}_{2,2}^A \left( \mathbf{M}_{1,2}^A \right)^{-1} \mathbf{M}_{1,1}^L - \mathbf{M}_{2,1}^L \right] \mathbf{L}^{(+)}$$
(B.14)

Equations (B.13) and (B.14) imply Equation (42).

**Appendix C**

In this appendix it is determined the states of total reflectance for the system of Fig. 5(a) with five equidistant terminals and $N_S = 10n$, being $n = 1, 2, 3, \cdots$, when all self-energies are zero and all hopping integrals are $t$.

This system has chains attached in the ring-sites $T_1 = 1$, $T_2 = 2n+1$, $T_3 = 4n+1$, $T_4 = 6n+1$ and $T_5 = 8n+1$. Notice that these are frontier-sites of the ring. Let us assume that there is an incoming wave through the chain attached to site 1. If we have a total reflection state, then the amplitude coefficients of the other attached chains should be zero, and in consequence, from equation (9) the amplitude of the coefficient in sites $T_2, T_3, T_4$ and $T_5$ is also zero, i.e.,

$$a_{T_i} = 0. \tag{C.1}$$

Moreover, equation (5) is rewritten in this instance as

$$t a_{T_i - 1} + t a_{T_i + 1} = 0, \tag{C.2}$$

which implies $a_{T_i - 1} = -a_{T_i + 1}$, being $i = 2, 3, 4$ and $5$. Solutions of equation (C.1) correspond to chains of $2n+1$ atoms with hopping integrals $t$ and null self-energies, where the amplitude of the coefficient of the first and the last atom is null. Those states have energies [16]

$$E = 2t \cos\left(\frac{m\pi}{2n}\right), \tag{C.3}$$

being $m = 1, 2, \cdots, 2n-1$, while the coefficient amplitudes between sites $T_i$ and $T_{i+1}$ are given by

$$a_{T_i + j}^{(m)} = \alpha_i^{(m)} \sin\left(\frac{mj\pi}{2n}\right), \tag{C.4}$$

with $j = 1, 2, \cdots, 2n-1$. Notice that equation (C.4) also implies that $a_{T_1} = 0$, which in terms of equation (9) indicates a phase change of $\pi$ between the incoming and outgoing waves. Note that

$$a_{T_i + 2n - 1}^{(m)} = \begin{cases} a_{T_i + 1}^{(m)} & \text{if } m \text{ odd} \\ -a_{T_i + 1}^{(m)} & \text{if } m \text{ even} \end{cases}. \tag{C.5}$$

Since the amplitude coefficients of the chain attached to $T_1$ are not zero, we require that

$$ta_2 + ta_{N_S} \neq 0. \tag{C.6}$$

Then, in order to accomplish Equation (C.2), (C.5) and (C.6), the energy (C.3) produces total reflection states if $m = 1, 3, 5, \cdots, 2n-1$.